\documentclass[twocolumn, switch]{article} 
\usepackage{preprint}
\usepackage{comment}
\usepackage{amsmath, amsthm, amssymb, amsfonts}

\usepackage[numbers,sort&compress]{natbib}
\usepackage[utf8]{inputenc}	
\usepackage[T1]{fontenc}	
\usepackage{xcolor}		
\usepackage[colorlinks = true,
            linkcolor = purple,
            urlcolor  = blue,
            citecolor = cyan,
            anchorcolor = black]{hyperref}	
\usepackage{booktabs} 		
\usepackage{nicefrac}		
\usepackage{microtype}		
\usepackage{lineno}		
\usepackage{float}			

\usepackage{lipsum}		
\usepackage{amsmath}
\usepackage{newfloat}
\DeclareFloatingEnvironment[name={Supplementary Figure}]{suppfigure}
\usepackage{sidecap}
\sidecaptionvpos{figure}{c}

\usepackage{titlesec}
\titlespacing\section{0pt}{10pt}{1pt}
\titlespacing\subsection{0pt}{10pt plus 3pt minus 3pt}{1pt plus 1pt minus 1pt}
\titlespacing\subsubsection{0pt}{8pt plus 3pt minus 3pt}{1pt plus 1pt minus 1pt}

\usepackage{tikz,xcolor,hyperref}

\definecolor{lime}{HTML}{A6CE39}

\title{Search for a Time-Dependent $Z'$ Resonance in the Dimuon Channel}

\usepackage{background}
\usepackage{xcolor}     

\backgroundsetup{
  scale=1,
  color=gray!90,
  opacity=1,
  angle=0,
  position=current page.south,
  vshift=1.2cm,
  contents={\texttt{*correspondence: marlon.brade@g.msuiit.edu.ph}}
}

\usepackage{authblk}

\author{Marlon P. Brade$^{1,2,3}$}
\author{Jeremiah D. Juevesano$^{2,3,4}$}
\author{Venus Abbegaile S. Carbonel$^{2,3}$}
\author{Karen E. Bustamante$^{2,3}$}
\author{Dennis C. Arogancia$^{2,3}$}
\author{Jan Mickelle V. Maratas$^{2,3}$}

\affil{$^{1}$Department of Physics,
Western Mindanao State University--Main,
Zamboanga City, Philippines}

\affil{$^{2}$Department of Physics,
Mindanao State University--Iligan Institute of Technology,
Iligan City, Philippines}

\affil{$^{3}$Mindanao Radiation Physics Center, Premier Research Institute of Science and Mathematics, MSU-Iligan Institute of Technology, Iligan City, Philippines}

\affil{$^{4}$Department of Mathematics, Physics, and Computer Science, University of the Philippines - Mindanao, Davao City, Philippines}

\makeatletter
\begin{document}
\twocolumn[ 
  \begin{@twocolumnfalse}
  
\maketitle

\begin{abstract}
We present a time-domain search strategy for resonances with periodically
varying masses in high-energy collider data. Conventional resonance
searches rely on time-integrated event samples and are therefore
insensitive to signals whose properties evolve during data taking. To
address this limitation, we develop a two-dimensional unbinned likelihood
framework in invariant mass and event time, allowing time-dependent
resonances to be identified through characteristic trajectories in the
$(m,t)$ plane rather than as stationary peaks in the invariant-mass
distribution.

As a benchmark scenario, we consider a $Z'$ boson arising from a gauged
$U(1)_{B-L}$ extension of the Standard Model and introduce a
phenomenological model in which the mediator mass undergoes periodic
temporal modulation. The resulting signal produces distinctive
correlations between invariant mass and event time that are not
accessible to conventional time-integrated analyses. The method is
applied to dimuon events from the CMS Open Data corresponding to the
Run~G data-taking period at $\sqrt{s}=13~\mathrm{TeV}$, with an
integrated luminosity of $7.54~\mathrm{fb}^{-1}$. No statistically
significant excess above the background-only hypothesis is observed for
any of the benchmark time-dependent signal scenarios considered.
Accordingly, 95\% confidence-level upper limits are derived on the
signal production cross section and the corresponding gauge coupling
$g_{Z'}$ as a function of the $Z'$ mass for several benchmark values of
the modulation amplitude. The resulting limits demonstrate that
incorporating temporal information can enhance sensitivity relative to
conventional time-integrated searches, particularly in the low- and
intermediate-mass regions where sufficient signal statistics are
available.
\end{abstract}
\vspace{0.35cm}

  \end{@twocolumnfalse} 
] 



\section{Introduction}

The Standard Model (SM) of particle physics has achieved remarkable success in describing fundamental particles and their interactions. Nevertheless, it does not account for several established phenomena, including the nature of dark matter, the origin of neutrino masses and oscillations, the hierarchy problem, and the matter--antimatter asymmetry of the Universe ~\cite{Hernandez:2025spl,Hindmarsh:2020hop, Balazs:2013cia}. Among these open questions, the nature of dark matter remains one of the central challenges in modern physics. Its existence is firmly established through multiple lines of astrophysical and cosmological evidence, including galactic rotation curves, gravitational lensing, large-scale structure formation, and precision measurements of the cosmic microwave background ~\cite{Trimble:1987ee}. However, its particle identity remains unknown, and no experiment to date has achieved a definitive direct or indirect detection.
A broad experimental program has been developed to probe possible interactions between dark matter and the SM, including underground direct-detection experiments, beam-dump and fixed-target facilities, astrophysical observations, and high-energy collider searches ~\cite{ALEPH:2013dgf,Cacciapaglia:2006,Das:2021esm,A1:2011yso,NA64:2019auh,NA64:2017vtt,KLOE-2:2016ydq,APEX:2011dww,Essig:2010xa,Freytsis:2009bh,BaBar:2014zli,BESIII:2017fwv,LHCb:2017trq}. In particular, collider and intensity-frontier experiments have explored a wide range of portal scenarios connecting dark matter to visible matter, yet no confirmed signal has been observed. This persistent absence of evidence motivates the exploration of alternatives beyond the traditional weakly interacting massive particle (WIMP) paradigm, including axions, sterile neutrinos, and particles residing in hidden or dark sectors~\cite{Wilczek:1977pj, Preskill:1982cy, Abbott:1982af, Dodelson:1993je, Arkani-Hamed:2008hhe}.

Among these possibilities, ultralight bosonic dark matter has attracted significant interest due to its distinctive wave-like behavior on macroscopic scales ~\cite{Kimball:2023vxk}. For sufficiently small masses, the large occupation number allows it to be described as a coherent classical field rather than a collection of individual particles. This field undergoes oscillations with a frequency set by its mass and can remain coherent over astrophysical timescales. Such behavior arises naturally in scenarios where the dark matter abundance is generated via the misalignment mechanism. A direct consequence is the possibility of inducing time-dependent effects in particle physics observables, leading to signatures that are not captured by conventional time-integrated searches. These time-varying phenomena provide a complementary avenue for probing dark matter interactions. The possibility that ultralight scalar dark matter induces time-dependent modulations of collider observables has been explored in previous studies~\cite{Guo:2022vxr, Bauer:2026dgu, Fieg:2026zdr}. In the present work, we consider the specific scenario in which the coherent oscillation of the scalar field induces a periodic modulation of the effective mediator mass, giving rise to characteristic time-dependent signatures at collider experiments.

In this work, we extend this framework to the case of a $Z'$ gauge boson associated with a gauged $U(1)'$ symmetry, in which the mediator couples directly to a conserved SM current. While the underlying mechanism of mass modulation remains similar, the phenomenology differs due to the distinct coupling structure and production channels. In particular, we focus on a $Z'$ boson arising from a gauged $U(1)_{B-L}$ extension of the SM, which provides a well-motivated benchmark ~\cite{Basso:2010pe, Basso:2008iv, Basso:2010jm}. A key feature of this scenario is that the signal does not appear as a stationary resonance in invariant mass, but instead evolves in time, tracing a trajectory in the $(m,t)$ plane. As a result, conventional analyses that rely on time-integrated invariant mass distributions may significantly dilute the signal, especially for large modulation amplitudes. To address this, we develop a time-domain search strategy based on a two-dimensional likelihood in invariant mass and time. This approach retains the correlation between these observables and allows for the direct reconstruction of the time-dependent signal.

Although the analysis is presented in the context of the $B-L$ model, the
methodology is formulated at the level of collider observables and is
therefore applicable to a broader class of scenarios involving
time-dependent mediator properties. 
The analysis is performed using dimuon events from the CMS Open Data
corresponding to an integrated luminosity of
$7.54\,\mathrm{fb}^{-1}$ collected during the 2016 Run~G data-taking
period~\cite{CMSOpenData_Run2016G_DoubleMuon,
CMSOpenData_Run2016G_DoubleMuon_NanoAOD}. Run~G spans approximately
$26.5$ days and provides event-by-event GPS timestamps, allowing the
likelihood to be constructed directly as a function of the event time.
The observed event-time distribution is incorporated explicitly through
the time-dependent normalization procedure described in Sec.~II, thereby accounting for the nonuniform detector exposure and
data-taking rate throughout the run.
Run~G was selected because it provides a substantial integrated
luminosity together with an observational window sufficiently long to
probe the benchmark oscillation periods considered in this work. 
Although the proposed methodology is, in principle, applicable to
multiple CMS data-taking periods, extending the analysis beyond a single
period introduces additional experimental and statistical
considerations. In particular, successive data-taking periods may be
separated by substantial intervals with no recorded collisions, even
though the hypothesized signal evolves continuously in time.

\subsection{\label{sec:level2}Model}

We consider an extension of the SM by an additional Abelian
$U(1)'$ gauge symmetry with gauge boson $Z'_\mu$. The present analysis is
formulated within a low-energy effective field theory in which the heavy
degrees of freedom responsible for the spontaneous breaking of the
$U(1)'$ symmetry have been integrated out. Consequently, the $Z'$ mass is
treated as an effective low-energy parameter originating from the heavy
symmetry-breaking sector, while the present work focuses on the collider
phenomenology arising from its time-dependent modulation. Although
renormalizable portal interactions between the heavy symmetry-breaking
scalar and the dark matter field are generally allowed, we consider the
regime in which such couplings are sufficiently suppressed that their
contributions to the low-energy phenomenology can be neglected.

The effective Lagrangian is given by
\begin{equation}
\mathcal{L}
=
-\frac{1}{4} Z'_{\mu\nu} Z'^{\mu\nu}
+\frac{1}{2}m_{0Z'}^{\,2}Z'_\mu Z'^{\mu}
+g_{Z'}Z'_\mu J^\mu_{Z'}\,,
\label{eq:LZp}
\end{equation}
where $Z'_{\mu\nu}=\partial_\mu Z'_\nu-\partial_\nu Z'_\mu$, $g_{Z'}$ denotes the $U(1)'$ gauge coupling, and $J^\mu_{Z'}$ is the corresponding fermionic current. The dark matter sector consists of a complex scalar field $\phi$
carrying $U(1)'$ charge $Q_\phi$. The ultralight scalar mass
$m_\phi$ is treated as an effective low-energy parameter, while its
ultraviolet origin and stabilization mechanism are assumed to arise from
physics above the effective theory. Accordingly, the present analysis
should be regarded as an effective description valid at the energy scales
relevant to collider phenomenology. The homogeneous dark matter field evolves according to
\begin{equation}
\ddot{\phi}
+
3H\dot{\phi}
+
m_\phi^{2}\phi
=
0\,,
\end{equation}
where $H$ is the Hubble expansion rate. At late times, the field undergoes coherent classical oscillations with
the approximate solution
\begin{equation}
\phi(t)
\simeq
\phi_0\cos(m_\phi t + \delta)
,
\end{equation}

\noindent where $\phi_0$ is the oscillation amplitude and $\delta$ is the phase of the oscillation. The interaction between the dark matter field and the $U(1)'$ gauge boson arises through the covariant derivative
\begin{equation}
D_\mu\phi
=
\left(
\partial_\mu
-
ig_{Z'}Q_\phi Z'_\mu
\right)\phi\,,
\end{equation}
which generates the interaction
\begin{equation}
|D_\mu\phi|^2
\supset
(g_{Z'}Q_\phi)^2
|\phi|^2
Z'_\mu Z'^{\mu}\,.
\end{equation}
Within the effective theory, the coherent oscillating dark matter background modifies the effective mass of the $Z'$ according to
\begin{equation}
m_{Z'}^{2}(t)
=
m_{0Z'}^{\,2}
+
(g_{Z'}Q_\phi)^2
|\phi(t)|^2\,.
\label{eq:mzp_general}
\end{equation}

Substituting the oscillatory solution into
Eq.~(\ref{eq:mzp_general}) gives
\begin{equation}
m_{Z'}^{2}(t)
=
m_{0Z'}^{\,2}
+
(g_{Z'}Q_\phi)^2
\phi_0^{\,2}
\cos^{2}(m_\phi t+\delta).
\label{eq:mzp_expand}
\end{equation}

Factoring out the constant mass scale yields
\begin{equation}
m_{Z'}^{2}(t)
=
m_{0Z'}^{\,2}
\left[
1+
\frac{(g_{Z'}Q_\phi)^2\phi_0^{\,2}}
{m_{0Z'}^{\,2}}
\cos^{2}(m_\phi t+\delta)
\right].
\end{equation}

It is therefore convenient to define the dimensionless modulation parameter
\begin{equation}
\kappa
\equiv
\frac{(g_{Z'}Q_\phi)^2\phi_0^{\,2}}
{m_{0Z'}^{\,2}},
\label{eq:kappa}
\end{equation}
which measures the amplitude of the dark-matter-induced mass modulation. The effective $Z'$ mass can then be written compactly as
\begin{equation}
m_{Z'}^{2}(t)
=
m_{0Z'}^{\,2}
\left(
1+\kappa\cos^{2}(m_\phi t+\delta)
\right).
\label{eq:mZpFinal}
\end{equation}

Equation~(\ref{eq:mZpFinal}) forms the basis of the phenomenological
analysis presented in this work. The four parameters
$\{m_{0Z'},\kappa,m_\phi,\delta\}$ completely specify the time-dependent
evolution of the resonance mass within the effective description
considered here.

\section{Methodology}

In this work, we develop a time-dependent search strategy for a neutral vector boson $Z'$ in scenarios where its effective mass is modulated by a coherently oscillating ultralight scalar background. This leads to a phenomenology that differs substantially from conventional resonance searches, where the signal is assumed to appear as a stationary peak in invariant mass distributions. In the present scenario, the resonance mass evolves in time, producing a signal with a non-trivial structure in the joint invariant mass--time plane. Consequently, analyses based solely on time-integrated invariant mass spectra can significantly dilute the signal, particularly when the modulation amplitude is large and the resonance sweeps across a broad mass range. To retain sensitivity to this behavior, we construct a two-dimensional likelihood in the observable space $(m,t)$, where $m \equiv m_{\mu\mu}$ is the dimuon invariant mass and $t$ is the event time. The signal probability density function (PDF) is explicitly defined in this two-dimensional space, allowing the correlation between invariant mass and time to be treated directly within the likelihood. Unlike standard bump hunts, where sensitivity arises from localized excesses in invariant mass, the signal considered here appears as a time-dependent trajectory in the $(m,t)$ plane. Preserving this structure is therefore essential, motivating the use of a fully unbinned likelihood that maximizes the information retained from the data.

\subsubsection{Signal and Background Probability Distribution Function}

\begin{figure}[!h]
\includegraphics[width=0.5\textwidth]{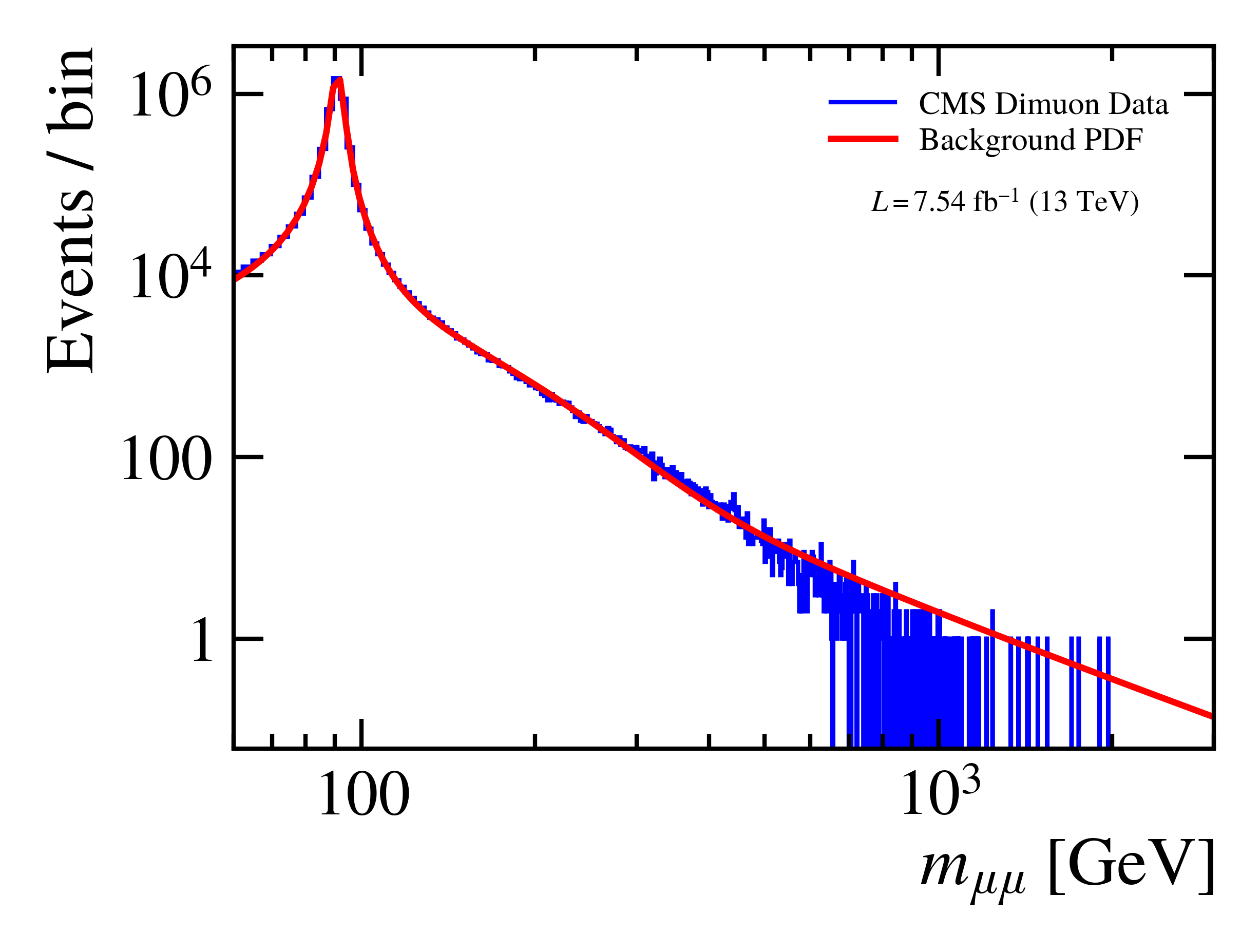}
\caption{Invariant mass distribution of dimuon events together with the fitted background model. The data are well described across the full mass range, including the $Z$-boson resonance and the smoothly falling high-mass tail. The fit is performed using an extended unbinned likelihood.}
\label{mass}
\end{figure}

\begin{figure}[!h]
\includegraphics[width=0.5\textwidth]{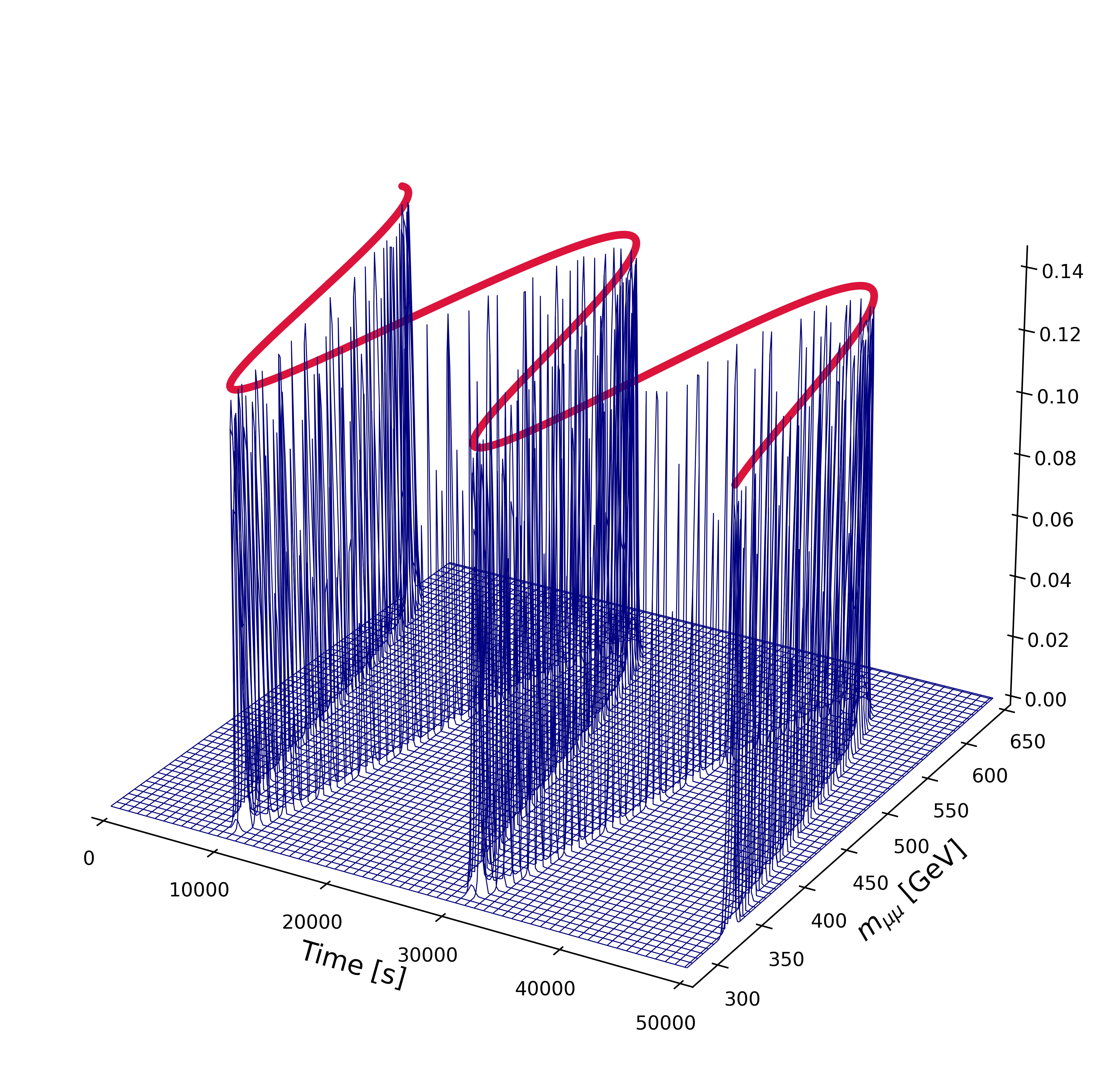}
\caption{Time-dependent evolution of the mass component of the signal PDF,
$S(m,t)$, for the benchmark point $m_0=300~\mathrm{GeV}$,
$\kappa=3$, and $\tau=20660~\mathrm{s}$. The Gaussian resonance follows
the time-dependent mean $\mu(t)$. The time-dependent acceptance function
$T(t)$ is applied separately in the full signal PDF used in the likelihood.}
\label{3d}
\end{figure}

\begin{figure}[!h]
\includegraphics[width=0.45\textwidth]{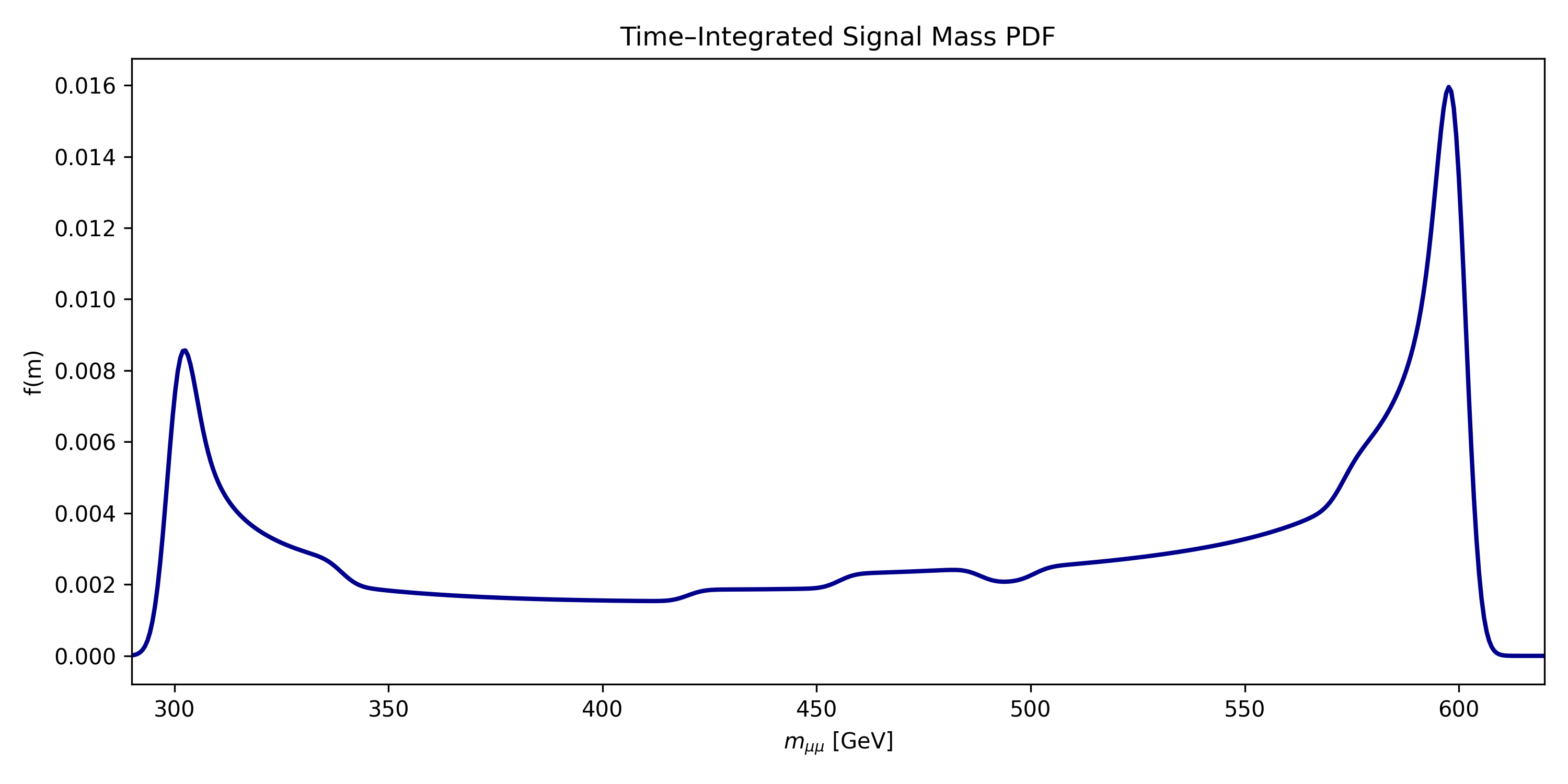}
\caption{Time-integrated signal probability density function obtained by integrating the two-dimensional signal template $S(m,t)$ over the full Run G data-taking period. The distribution is shown for the benchmark point $m_0=300~\mathrm{GeV}$ and $\kappa=3$. }
\label{smass}
\end{figure}

The analysis is performed in the two-dimensional observable space defined by the invariant mass $m$ and event time $t$. Figure~\ref{mass} shows the invariant mass distribution of the selected dimuon events together with the fitted background model. Good agreement is observed over the full mass range considered in the analysis, including both the resonant $Z$-boson peak and the non-resonant Drell--Yan continuum at higher masses.

The signal PDF is constructed to describe the time-dependent evolution of
the $Z'$ resonance mass induced by the oscillating scalar-field dark matter
background. At a fixed time, the reconstructed invariant-mass distribution
is modeled by a Gaussian function,
\begin{equation}
S(m,t)=
\frac{1}{\sqrt{2\pi}\,\sigma_{\mathrm{res}}}
\exp\!\left[
-\frac{\left(m-\mu(t)\right)^2}
{2\sigma_{\mathrm{res}}^2}
\right],
\end{equation}
where $\sigma_{\mathrm{res}}$ denotes the detector mass resolution.
Throughout this analysis, the resolution is parameterized as
$\sigma_{\mathrm{res}}=0.01\,m_0$. For the low-mass region,
$50\leq m_0\leq120~\mathrm{GeV}$, a value of
$\sigma_{\mathrm{res}}=0.02\,m_0$ is adopted. In the vicinity of the
$Z$-boson resonance, the large Standard Model background, together with a
narrow signal parameterization, makes the evaluation of the expected
upper limits numerically challenging. The broader signal
parameterization is therefore adopted only in this mass region to ensure
a stable likelihood evaluation.

The time-dependent mean of the resonance is parameterized as
\begin{equation}
\mu(t)
=
m_0
\sqrt{
1+\kappa
\cos^2
\left(
\frac{\pi t}{\tau}+\delta
\right)
},
\label{eq:mu}
\end{equation}
where $m_0$ is the nominal resonance mass, $\kappa$ controls the modulation
amplitude, $\tau$ is the oscillation period, and $\delta$ is an arbitrary
phase offset. This parameterization provides a minimal phenomenological
description of a resonance whose mass undergoes periodic modulation due to
the coherent oscillation of the scalar dark matter field. Consequently,
the signal exhibits a non-factorizable correlation between invariant mass
and event time, which constitutes the characteristic experimental
signature exploited in the present analysis.

Throughout this work, we consider a benchmark scalar mass of
$m_\phi \simeq 10^{-19}\,\mathrm{eV}$,
corresponding to an oscillation period of
$\tau \simeq 20660~\mathrm{s}$
($\approx 5.7$ hours). This benchmark is particularly well suited for the
present analysis because the modulation is neither sufficiently rapid to
be washed out by the finite event statistics nor sufficiently slow to
sample only a small fraction of an oscillation cycle during the data-taking
period. Instead, multiple oscillation cycles are expected to occur over the
CMS Run~G dataset while retaining adequate statistics within each cycle to
reconstruct the evolving resonance structure. Furthermore, the coherence
time of an ultralight scalar field with this mass is expected to greatly
exceed the duration of the dataset, allowing the modulation to be treated
as a coherent periodic signal throughout the analysis. To account for variations in detector exposure throughout the
data-taking period, the signal probability density function is
multiplied by a time-dependent acceptance function $T(t)$,
\begin{equation}
P_{\mathrm{sig}}(m,t)
=
S(m,t)\,T(t).
\end{equation}

The background contribution is modeled as the sum of two components
corresponding to resonant $Z$-boson production and the non-resonant
Drell--Yan continuum,
\begin{equation}
P_{\mathrm{bkg}}(m,t)
=
P_Z(m,t)
+
P_{\mathrm{DY}}(m,t).
\end{equation}

\noindent The resonant $Z$-boson contribution is modeled using a Double Crystal
Ball (DCB) function, which describes the Gaussian core together with
asymmetric non-Gaussian tails arising from detector resolution effects
and final-state radiation. The DCB shape parameters are fixed using CMS
Open Data simulation samples \cite{CMSOpenData_ZToMuMu_M50To120_NanoAODSIM}. The non-resonant Drell--Yan contribution
is modeled using a log-curved power-law function,
\begin{equation}
B_{\mathrm{DY}}(m)
\propto
m^{-(n+b\log m)},
\end{equation}
which provides a flexible parametrization of the smoothly falling
invariant-mass spectrum over the full search region.

\begin{figure*}[!h]
\centering
\includegraphics[width=1.0\textwidth]{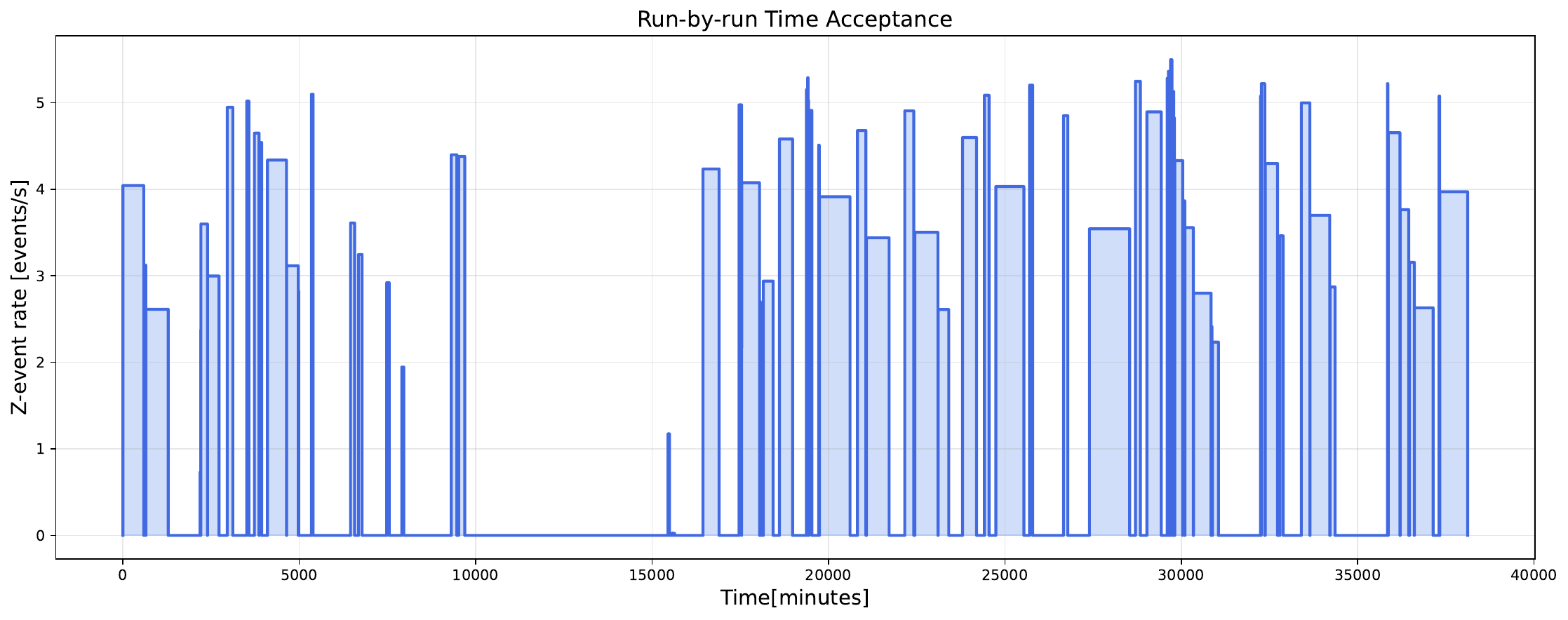}
\caption{Piecewise-constant time-dependent acceptance function derived
from the CMS Run~G Open Data. For each CMS run, the acceptance is
computed as $N_Z/\Delta t$, where $N_Z$ is the number of reconstructed
$Z\rightarrow\mu^+\mu^-$ candidates satisfying
$80<m_{\mu\mu}<100~\mathrm{GeV}$ and $\Delta t$ is the duration of the
corresponding run. The resulting function provides a data-driven
estimate of the effective detector exposure, incorporating the combined
effects of the average luminosity, detector acceptance, and
data-taking efficiency, and is applied as the common time-dependent
acceptance function for both the signal and background probability
density functions.}
\label{time}
\end{figure*}

Both background components are multiplied by the common time-dependent
acceptance function $T(t)$, shown in Fig.~\ref{time}. The acceptance
function is constructed from individual CMS run numbers, where each run
corresponds to a contiguous data-taking period during which the detector
configuration, calibration constants, trigger menu, and operating
conditions remain approximately stable. For each run $i$, spanning the
time interval $[t_i^{\min},t_i^{\max})$, the acceptance is assigned the
value
\begin{equation}
T(t)
=
\frac{N_Z^{(i)}}{\Delta t_i},
\qquad
t\in[t_i^{\min},t_i^{\max}),
\end{equation}
where $N_Z^{(i)}$ denotes the number of reconstructed
$Z\rightarrow\mu^+\mu^-$ candidates satisfying
\begin{equation}
80~\mathrm{GeV}
<
m_{\mu\mu}
<
100~\mathrm{GeV},
\end{equation}
and
\begin{equation}
\Delta t_i
=
t_i^{\max}-t_i^{\min}
\end{equation}
is the duration of the corresponding CMS run. Since the inclusive $Z$-boson production cross section is expected to
remain constant during data taking, the observed $Z$-boson event rate
provides a data-driven estimate of the effective detector exposure.
Consequently, the acceptance function incorporates the combined effects
of the average luminosity during each CMS run, detector acceptance, and
data-taking efficiency, thereby providing a piecewise-constant
description of the slowly varying detector exposure throughout the
data-taking period.

The run-based construction of the acceptance function represents a
physically motivated compromise between detector stability and
statistical precision. Because detector configuration, calibration,
trigger settings, and operating conditions are expected to remain
approximately stable within a given CMS run, constructing the acceptance
function on a run-by-run basis naturally preserves these periods of
stable detector operation. In contrast, arbitrary fixed-width time
intervals may span multiple CMS runs, thereby mixing data collected
under different detector configurations or operating conditions into a
single acceptance estimate. Conversely, substantially finer intervals,
such as individual luminosity sections, contain significantly fewer
$Z$-boson events and consequently lead to larger statistical
fluctuations in the estimated detector exposure. Each CMS run therefore
provides sufficient $Z$-boson statistics to obtain a stable and
physically meaningful estimate of the relative detector exposure.

The acceptance function serves only as a normalization correction and
does not determine the time resolution of the signal model. For each
event, the signal probability density function is evaluated using the
exact event timestamp through the time-dependent mean $\mu(t)$ defined
in Eq.~(\ref{eq:mu}), while the corresponding value of $T(t)$ is
determined solely by the CMS run in which the event was recorded.
Consequently, the temporal evolution of the signal is treated
continuously on an event-by-event basis, whereas the acceptance function
corrects only for the slowly varying detector exposure and does not
affect the continuous time evolution described by
Eq.~(\ref{eq:mu}).

The background probability density functions are therefore written as
\begin{equation}
\begin{aligned}
P_Z(m,t)
&=
Z(m)\,T(t),
\\
P_{\mathrm{DY}}(m,t)
&=
B_{\mathrm{DY}}(m)\,T(t).
\end{aligned}
\end{equation}

\begin{figure*}[t]
\centering
\includegraphics[width=\textwidth]{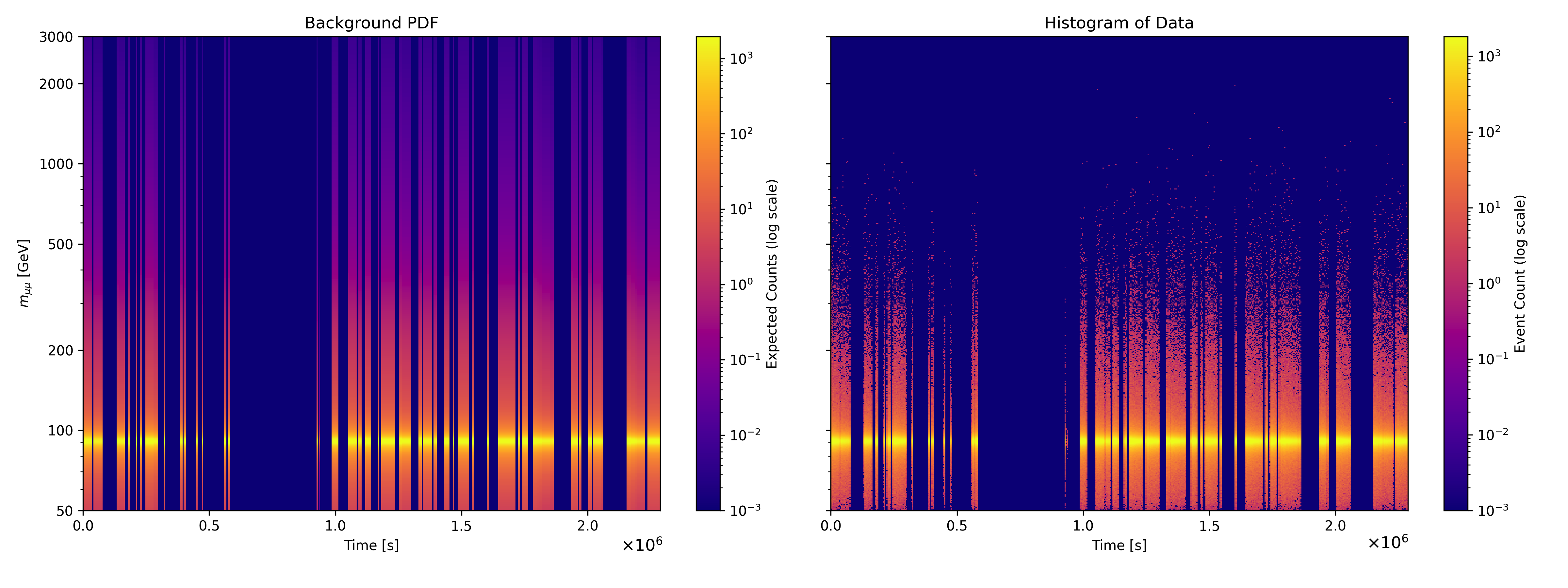}
\caption{Illustration of the background structure in the $(m,t)$ plane.}
\label{heatmap}
\end{figure*}

The complete likelihood is constructed as an extended model in which the
signal and background yields are determined simultaneously from the fit.
While the SM background factorizes into independent
invariant-mass and time components through the common acceptance
function, the signal remains intrinsically non-factorizable owing to the
time-dependent mean defined in Eq.~(\ref{eq:mu}). This correlation
between invariant mass and event time constitutes the principal
discriminating feature of the analysis, enabling a periodically varying
resonance to be distinguished from SM backgrounds. The
overall structure of the background model together with the observed
event distribution in the $(m,t)$ plane is illustrated in
Fig.~\ref{heatmap}.

\subsection{Statistical Analysis}
\label{sec:8}

Statistical inference is performed using a modified frequentist approach
based on the profile likelihood method~\cite{Cowan:2010js,Moneta:2010pm}.
The analysis employs an extended unbinned likelihood constructed in the
two-dimensional observable space of invariant mass and event time,
thereby preserving the full event-by-event information while comparing
the signal and background hypotheses.
The parameter of interest is the signal yield, $N_{\rm sig}$. The
background normalizations, $N_Z$ and $N_{\rm pow}$, are treated as
nuisance parameters and are profiled during the likelihood
maximization. All remaining model parameters are determined through
auxiliary studies prior to the final hypothesis test and are
subsequently fixed. The tail parameters of the Double Crystal Ball function
($\alpha_L$, $n_L$, $\alpha_R$, and $n_R$) are first determined from CMS
Open Data simulation. These parameters are then fixed in a preliminary
fit to the invariant-mass distribution of the data, in which the
$Z$-boson peak position ($\mu_Z$), detector mass resolution
($\sigma_Z$), the Drell--Yan shape parameters
($n_{\rm pow}$ and $b_{\rm pow}$), together with the background
normalizations ($N_Z$ and $N_{\rm pow}$), are allowed to vary. The
resulting best-fit values define the background model used in the
subsequent statistical analysis.

For each benchmark signal hypothesis, a preliminary two-dimensional fit
is subsequently performed to determine the best-fit value of the signal
phase $\delta$. The fitted value of $\delta$, together with the
calibrated background shape parameters and the benchmark signal
parameters ($m_0$, $\kappa$, and $\tau$), is then fixed in the final
profile likelihood. Consequently, the final statistical inference
contains only three free parameters: the parameter of interest,
$N_{\rm sig}$, and the nuisance parameters, $N_Z$ and
$N_{\rm pow}$. This reduces the dimensionality of the likelihood
optimization while preserving the independently calibrated signal and
background models. The profile likelihood ratio test statistic is defined as

\begin{equation}
q_\mu
=
-2
\ln
\left(
\frac{\mathcal{L}(\mu,\hat{\hat{\theta}})}
{\mathcal{L}(\hat{\mu},\hat{\theta})}
\right),
\end{equation}

\noindent where $\mu$ denotes the signal strength parameter,
$\hat{\mu}$ and $\hat{\theta}$ are the values of the signal strength and
nuisance parameters that globally maximize the likelihood function, and
$\hat{\hat{\theta}}$ denotes the conditional maximum-likelihood
estimator of the nuisance parameters for a fixed value of $\mu$.

In the absence of a statistically significant deviation from the
background-only expectation, upper limits are derived using the
$CL_s$ prescription. Exclusion limits at the 95\% confidence level are
obtained for the signal hypothesis parameterized by $\mu$, which in the
present analysis corresponds to the product
$\sigma(pp\rightarrow Z')\times
\mathrm{BR}(Z'\rightarrow\ell^+\ell^-)$. Signal hypotheses satisfying
$CL_s<0.05$ are considered excluded at the 95\% confidence level.

\subsection{Determination of the \texorpdfstring{$(M_{Z'},g_{Z'})$}{(MZ', g')} Exclusion Limits}
\label{sec:9}

The exclusion limits in the $(M_{Z'},g_{Z'})$ parameter space are obtained by translating the expected 95\% confidence level upper limits on the signal production cross section into constraints on the gauge coupling. For each benchmark mediator mass, the upper limit on $\sigma(pp\rightarrow Z')\times\mathrm{BR}(Z'\rightarrow\ell^+\ell^-)$ is first determined using the $\mathrm{CL_s}$ procedure described in the previous section.

To convert these cross-section limits into limits on the coupling, Monte Carlo samples are generated for each mediator mass using \textsc{MadGraph5\_aMC@NLO} ~\cite{Alwall:2014hca}. The same generator-level requirements adopted in the signal efficiency determination are applied to ensure consistency between the simulated signal acceptance and the efficiency used in the statistical analysis. The coupling $g_{Z'}$ is then varied while keeping the mediator mass fixed until the predicted production cross section multiplied by the dilepton branching ratio matches the corresponding expected upper limit obtained from the $\mathrm{CL_s}$ analysis. Repeating this procedure for all considered mass hypotheses yields the exclusion contour in the $(M_{Z'},g_{Z'})$ plane.

\section{Results and discussion}
\label{R&D}

\begin{figure}[!h]
\includegraphics[width=0.45\textwidth]{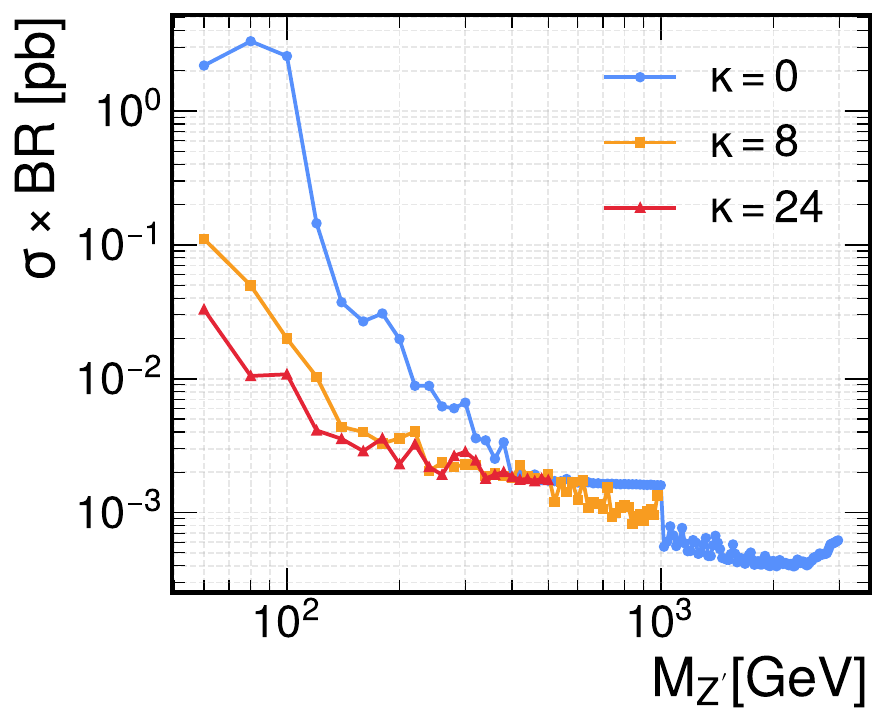}
\caption{Expected cross section times branching ratio as a function of the mediator mass for different values of the modulation parameter $\kappa$. The signal rate decreases rapidly with increasing mass.}
\label{cross_section}
\end{figure}

The analysis is performed using the CMS Open Data corresponding to the Run G
period of the 2016 proton--proton collision dataset at
$\sqrt{s}=13~\mathrm{TeV}$. Events are selected from the
\texttt{DoubleMuon} primary dataset and are required to satisfy at least one
double-muon High-Level Trigger (HLT). The trigger menu consists of dimuon
triggers requiring two reconstructed muons with transverse momentum
thresholds of 17 and 8~GeV, including standard and tracker-muon
configurations, variants requiring compatibility with a common longitudinal
interaction vertex ($\Delta z$), and isolated dimuon triggers for the
leading muon within $|\eta|<2.1$. Since the offline analysis requires both
muons to satisfy $p_T>25~\mathrm{GeV}$, the selected events are well within
the trigger efficiency plateau. For each event, a unique dimuon candidate is constructed from the
reconstructed muons. The muon with the highest transverse momentum
satisfying $p_T>25~\mathrm{GeV}$ and $|\eta|<2.4$ is identified as the
leading muon. Among the remaining reconstructed muons, the highest-$p_T$
muon with opposite electric charge satisfying the same kinematic
requirements is selected as the second muon. Events for which no such
opposite-sign partner exists are rejected. Consequently, exactly one
dimuon candidate is retained per event, avoiding multiple entries from
events containing more than two reconstructed muons.

The selected dimuon candidate is subsequently required to satisfy an
invariant mass of $m_{\mu\mu}>50~\mathrm{GeV}$. To ensure a high-purity
sample of promptly produced muons, both muons are required to satisfy
either the CMS Medium or Tight identification working point, together with
the impact parameter requirements
$|d_{xy}|<0.05~\mathrm{cm}$ and
$|d_z|<0.10~\mathrm{cm}$.
Backgrounds from nonprompt muons originating from heavy-flavor decays and
hadronic activity are further suppressed by requiring both muons to satisfy
the particle-flow relative isolation criterion
$I_{\mathrm{rel}}^{0.4}<0.15$.
The complete set of offline event selection
requirements, including the kinematic acceptance, invariant mass
requirement, muon identification criteria, impact parameter cuts,
isolation requirement, and transverse momentum threshold, are summarized in
Table~\ref{tab:event_selection}.

\begin{table*}[t]
\centering
\caption{Summary of the trigger requirements and offline event selection used in the analysis.}
\label{tab:event_selection}
\renewcommand{\arraystretch}{1.25}
\begin{tabular}{lp{11cm}}
\hline\hline
\textbf{Category} & \textbf{Requirement} \\
\hline

Invariant Mass &
$m_{\mu\mu}>50~\mathrm{GeV}$ \\

Muon Identification &
Both muons satisfy the CMS Medium or Tight identification working point. \\

Muon Acceptance &
Both muons satisfy $|\eta|<2.4$. \\

Transverse Momentum &
Both muons satisfy $p_T>25~\mathrm{GeV}$. \\

Transverse Impact Parameter &
Both muons satisfy $|d_{xy}|<0.05~\mathrm{cm}$. \\

Longitudinal Impact Parameter &
Both muons satisfy $|d_z|<0.10~\mathrm{cm}$. \\

Muon Isolation &
Both muons satisfy the particle-flow relative isolation requirement
$I_{\mathrm{rel}}^{0.4}<0.15$. \\

\hline\hline
\end{tabular}
\end{table*}

The two-dimensional event distribution in the $(m,t)$ plane exhibits the
expected structure associated with the time-dependent normalization
function incorporated into the likelihood model. As shown in
Fig.~\ref{heatmap}, the observed variations in event density primarily
reflect changes in the data-taking conditions, including variations in
luminosity and run duration, rather than genuine physical modulations.
These temporal features are well reproduced by the background model,
demonstrating that the normalization function provides an adequate
description of the time dependence of the dataset. The excellent
agreement between the observed data and the background prediction
therefore validates the treatment of detector acceptance and
data-taking conditions adopted in the time-dependent likelihood
analysis.

As shown in Fig.~\ref{cross_section}, the expected signal production rate
decreases rapidly with increasing mediator mass owing to the falling
production cross section. Consequently, the number of expected signal
events decreases, reducing the statistical sensitivity of the search in
the high-mass region. This trend is reflected in the expected upper
limits on $\sigma\times\mathrm{BR}$ for all values of $\kappa$,
although the time-dependent scenarios consistently retain improved
sensitivity relative to the time-independent case ($\kappa=0$).

The observed improvement in sensitivity can be understood from the
properties of the signal model. As illustrated in Fig.~\ref{smass}, a
nonzero modulation amplitude ($\kappa>0$) gives rise to a characteristic
double-peaked time-integrated signal distribution. As the modulation
amplitude increases, the separation between the two peaks becomes
larger, shifting the upper peak toward higher invariant masses where the
smoothly falling Drell--Yan background is substantially reduced.
Consequently, a larger fraction of the signal populates a region with a
more favorable signal-to-background ratio. At the same time, larger
values of $\kappa$ produce a more pronounced time-dependent structure in
the two-dimensional $(m,t)$ space, providing additional discriminating
information beyond that available from the invariant-mass distribution
alone. The two-dimensional likelihood is therefore able to exploit both
the modified signal shape and the temporal information to improve the
separation between the signal and the Standard Model background,
resulting in enhanced sensitivity without requiring an increase in the
total number of signal events.

\begin{figure}[!h]
\includegraphics[width=0.45\textwidth]{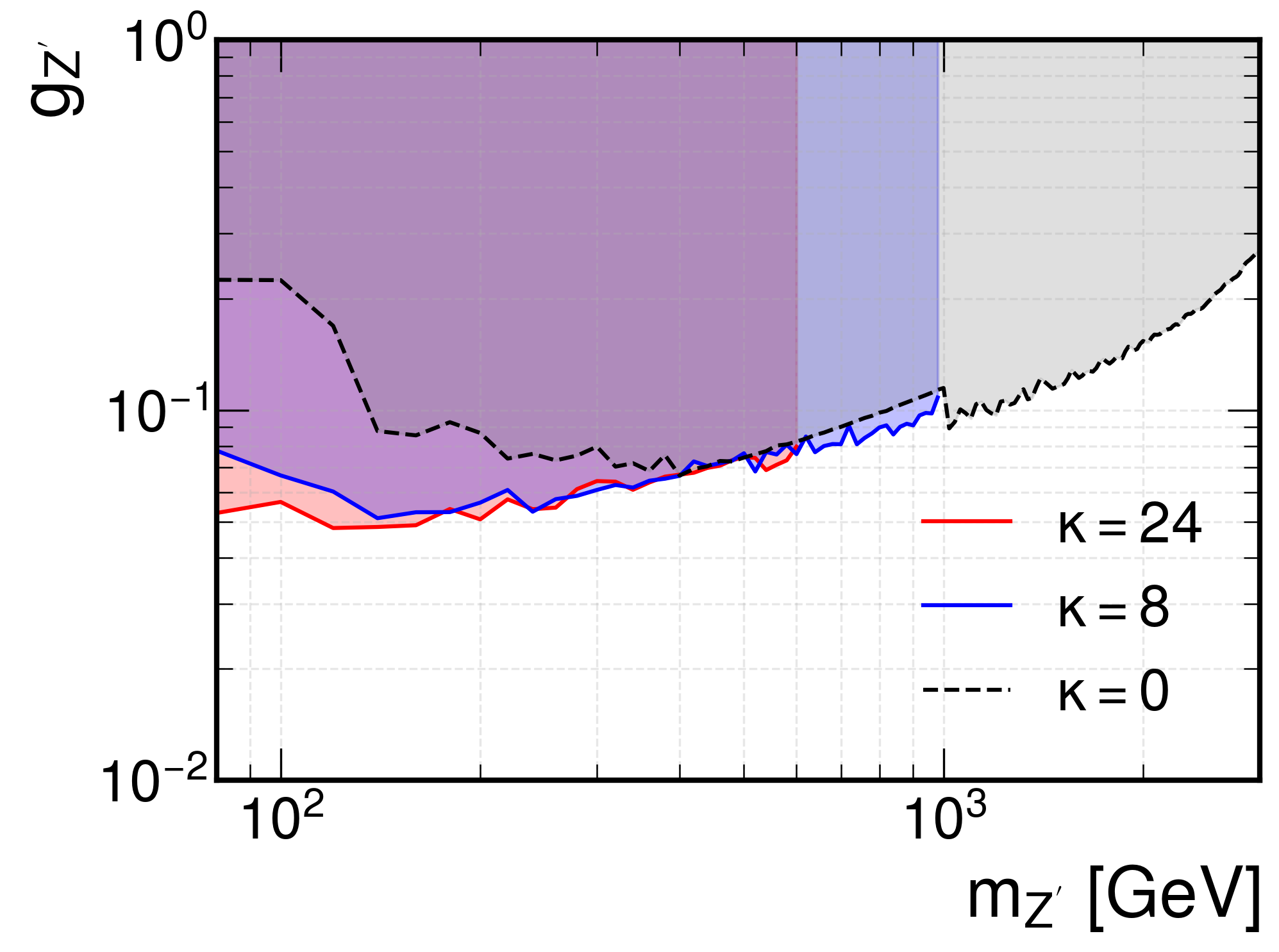}
\caption{Upper limits on the coupling $g_{Z'}$ as a function of the mediator mass $m_{Z'}$ for different values of the modulation parameter $\kappa$. The dashed curve corresponds to the time-independent case ($\kappa = 0$), while solid curves show the time-dependent results.}
\label{coup}
\end{figure}

The agreement between the observed data and the background model in the
$(m,t)$ plane indicates that no statistically significant deviation from
the SM expectation is observed for any of the benchmark
time-dependent signal hypotheses considered in this work.
Consequently, the data are found to be consistent with the
background-only hypothesis, and 95\% confidence level upper limits are
derived on the signal production cross section times branching ratio and
the corresponding gauge coupling. The resulting exclusion limits on
$g_{Z'}$ are shown in Fig.~\ref{coup}, where the time-independent scenario
($\kappa=0$) is represented by the dashed curve and the time-dependent
cases with $\kappa=8$ and $\kappa=24$ by the solid curves. A clear and
systematic improvement in sensitivity is observed as the modulation
amplitude increases, allowing progressively smaller couplings to be
excluded for a fixed mediator mass. At higher mediator masses, the
limits become weaker for all values of $\kappa$, primarily because the
rapidly decreasing production cross section reduces the available signal
statistics. Consequently, the separation between the different
$\kappa$ scenarios gradually decreases in the high-mass region.
Nevertheless, even with the limited integrated luminosity of the CMS
Run~G Open Data, the present analysis demonstrates that time-domain
resonance searches using event-by-event timing information are
experimentally feasible. The sensitivity of this approach is expected to
improve with larger datasets, particularly in the high-mass region where
the current reach is primarily limited by statistical precision.

\section{Summary and Conclusions}
\label{SC}

In this work, we have developed a time-dependent search strategy for vector mediators whose masses are modulated by a coherently oscillating ultralight scalar dark matter field. As a benchmark scenario, we considered a $Z'$ gauge boson arising from a gauged $U(1)_{B-L}$ extension of the SM, in which the interaction with the scalar field induces a periodic variation of the effective mediator mass. Unlike conventional resonance searches that rely solely on the invariant-mass spectrum, such a signal produces a characteristic trajectory in the joint invariant mass--time $(m,t)$ space, providing an additional observable that can be exploited in the statistical analysis.

To capture this distinctive signature, we constructed a two-dimensional extended unbinned likelihood in invariant mass and event time and applied the method to dimuon events from the CMS Run~G Open Data corresponding to an integrated luminosity of $7.54~\mathrm{fb}^{-1}$ at $\sqrt{s}=13~\mathrm{TeV}$. A data-driven background model, together with a time-dependent acceptance function derived from the observed $Z$-boson yield, was used to account for variations in detector conditions and instantaneous luminosity throughout the data-taking period. No statistically significant deviation from the SM expectation was observed, and upper limits at the 95\% confidence level were established on the production cross section times branching ratio and the corresponding gauge coupling.

The resulting exclusion limits demonstrate that incorporating temporal
information systematically improves the sensitivity relative to the
time-independent scenario ($\kappa=0$). This improvement can be
understood from the properties of the signal model. As the modulation
amplitude increases, the time-integrated signal distribution develops
enhanced probability density near the turning points of the oscillation,
shifting a larger fraction of the signal toward higher invariant masses
where the smoothly falling Drell--Yan background provides a more
favorable signal-to-background ratio. At the same time, the periodic
modulation gives rise to a characteristic correlation between invariant
mass and event time that is absent in the Standard Model background,
providing additional discriminating information beyond that contained in
the invariant-mass distribution alone. Consequently, the greatest
improvement is achieved in the low- and intermediate-mass regions,
whereas at higher masses the sensitivity becomes increasingly limited by
the rapidly decreasing signal yield and the corresponding statistical
uncertainty.

Overall, this work establishes a general framework for incorporating event time as an additional observable in collider searches for time-dependent new physics. Although demonstrated here for a time-varying $Z'$ boson using a subset of CMS Open Data, the methodology is formulated at the level of collider observables and is readily applicable to a broader class of scenarios involving time-dependent mediator properties. Future analyses using larger datasets and higher integrated luminosities are expected to further extend the discovery potential of this approach.

\section*{Acknowledgments}
The authors would like to acknowledge the Department of Science and Technology--Advanced Science and Technology Institute (DOST--ASTI) for providing access to the High-Performance Computing (HPC) resources used in this work. We also gratefully acknowledge the Department of Science and Technology--Accelerated Science and Technology Human Resource Development Program (DOST--ASTHRDP) for scholarship support.

The authors sincerely thank Dr.~Camille Normand and Dr.~Kostantinos Petridis of the University of Bristol for their guidance and valuable discussions during the development and implementation of the time-dependent analysis methodology that formed the basis of this work. In addition, we acknowledge the CMS Collaboration for making the CMS Open Data publicly available, which enabled this study.

\normalsize
\bibliography{references}


\end{document}